# The Health–Mortality Approach in Estimating the Healthy Life Years Lost Compared to the Global Burden of Disease Studies and Applications in World, USA and Japan*

Christos H Skiadas

ManLab, Technical University of Crete, Chania, Crete, Greece
Email: skiadas@cmsim.net

**Abstract**

We propose a series of methods and models in order to explore the Global Burden of Disease Study and the provided healthy life expectancy (HALE) estimates from the World Health Organization (WHO) based on the mortality $\mu_x$ of a population provided in a classical life table and a mortality diagram. Our estimates are compared with the HALE estimates for the World territories and the WHO regions along with providing comparative results with to findings of Chang, Molla, Truman et al. (2015) on the "Differences in healthy life expectancy for the US population by sex, race/ethnicity and geographic region: 2008 for USA" and from Yong and Saito (2009) regarding "Trends in healthy life expectancy in Japan".

From the mortality point of view we have developed a simple model for the estimation of a characteristic parameter *b* related to the healthy life years lost to disability and providing full application details along with characteristic parameter selection and stability of the coefficients. We also provide a direct estimation method of the parameter *b* from the life tables. We straighten the importance of our methodology by proposing and applying estimates of the parameter *b* by using the Gompertz and the Weibull models.

From the Health State point of view we summarize the main points of the first exit time theory to life table data and present the basic models. Even more we develop the simpler 2-parameter health state model and an extension of a model expressing the infant mortality to a 4-parameter model which is the simpler model providing very good fitting on the logarithm of the force of mortality, $\ln(\mu_x)$. More important is the use of the Health State Function and the relative impact on mortality to find an estimate for the healthy life years lost to disability.

We have developed simple programs in Excel providing immediately the Life Expectancy, the Loss of Healthy Life Years and the Healthy Life Expectancy estimate.

## 1. Introduction

Starting from the late 80's a Global Burden of Disease (GBD) study was applied in many countries reflecting the optimistic views of many researchers and policy makers worldwide to quantify the health state of a population or a group of persons. In the time course they succeeded in establishing an international network collecting and providing adequate information to calculate health measures under terms as Loss of Healthy Life Years (LHLY) or Healthy Life Expectancy (HALE). The latter tends to be a serious measure important for the policy makers and national and international health programs. So far the process followed was towards statistical measures including surveys and data collection using questionnaires and disability and epidemiological data as well (McDowell, 2006). They faced many views referring to the definition of health and to the inability to count the various health states and of course the different cultural and societal aspects of the estimation of health by various persons worldwide. Further to any objections posed when trying to quantify health, the scientific community had simply to express with strong and reliable measures that millions of people for centuries and thousands of years expressed and continue to repeat every day: That their health is good, fair, bad or very bad. As for many decades the public opinion is seriously quantified by using well established statistical and poll techniques it is not surprising that a part of these achievements helped to improve, establish and disseminate the health state measures. However, a serious scientific part is missing or it is not very much explored that is to find the model underlying the health state measures. Observing the health state measures by country from 1990 until nowadays it is clear that the observed and estimated health parameters follow a rather systematic way. If so why not to find the process underlying these measures? It will support the provided health measures with enough documentation while new horizons will open towards better estimates and data validation.

___
*Paper version–submitted to ArXiv.org (8 March 2016)

From the early 90's we have introduced and applied methods, models and techniques to estimate the health state of a population. The related results appear in several publications and we have already observed that our estimates are related or closely related to the provided by the World Health Organization (WHO) and other agencies as Eurostat or experts as the REVES group. However, our method based on a difficult stochastic analysis technique, is not easy to use especially by practitioners. The last four centuries demography and demographers are based on the classical Life Tables. Thus here we propose a very simple model based on the mortality $\mu_x$ of a population provided in a classical life table. To compare our results with those provided by WHO we use the $\mu_x$ included in the WHO abridged life tables. Our estimates are compared with the HALE estimates for all the WHO countries. Even more we provide the related simple program in Excel which provides immediately the Life Expectancy, the Loss of Healthy Life Years and the Healthy Life Expectancy estimate. The comparisons suggest an improved WHO estimate for the majority of the countries. There are countries' results differing from the model and need further study.

**Further Details**

The Global Burden of Disease Study explored the health status of the population of all the countries members of the World Health Organization (WHO). It is a large team work started more than 25 years ago (see Murray and Lopez, 1997,2000, Mathers et al., 2000, Salomon, et al., 2010, 2012, Murray et al., 2015, Hausman, 2012, Vos et al., 2012, Robine, Romieu, Cambois, 1999, WHO, 2000, 2001, 2002, 2004, 2013, 2014 and many other publications). The last years, with the financial support of the Bill and Melinda Gates foundation, the work was expanded via a large international group of researchers. The accuracy of the data collection methods was improved along with the data development and application techniques. So far the health status indicators were developed and gradually were established under terms as healthy life expectancy and loss of healthy life years. Methods and techniques developed during the seventies and eighties as the Sullivan method (Sullivan, 1971) were used quite successfully. Several publications are done with the most important included in The Lancet under the terms DALE and HALE whereas a considerable number can be found in the WHO and World Bank publications. The same half part of a century several works appear in the European Union exploring the same phenomenon and providing more insight to the estimation of the health state of a population and providing tools for the estimation of severe, moderate and light disability. The use of these estimates from the health systems and the governments is obvious.

To a surprise the development of the theoretical tools was not so large. The main direction was towards to surveys and collection of mass health state data instead of developing and using theoretical tools. The lessons learned during the last centuries were towards the introduction of models in the analysis of health and mortality. The classical examples are Edmund Halley for Life Tables and Benjamin Gompertz for the law of mortality and may others. Today our ability to use mass storage tools as the computers and the extensive application of surveys and polls to many political, social and economic activities directed the main health state studies. In other words we give much attention to opinions of the people for their health status followed by extensive health data collection. However, it remains a serious question: can we validate the health status results? As it is the standard procedure in science a systematic study as the Global Burden of Disease should be validated by one or more models. Especially as these studies are today the main tool for the health programs of many countries the need of verification is more important.

People reply according to their experience. Two main approaches arise: The mortality focus approach and the health status approach. Although both look similar responds may have significant differences. The main reason is that health is a rather optimistic word opposed to the pessimistic mortality term. Twenty years ago we provided a model to express the health state of a population. We developed and expanded this model leading to a system providing health status indexes. Here we propose several methodologies to estimate the health indexes and to compare with the provided by WHO.

**2. The mortality approach**

**2.1. The Simplest Model**

We need a simple model to express the health status. The best achievement should be to propose a model in which the health measure should be presented by only one main parameter. We thus propose a two parameter model with one crucial health parameter:

$$\mu_x = \left(\frac{x}{T}\right)^b \tag{1}$$

The parameter *T* represents the age at which $\mu_x$=1 and *b* is a crucial health state parameter expressing the curvature of $\mu_x$. As the health state is improved *b* gets higher values.

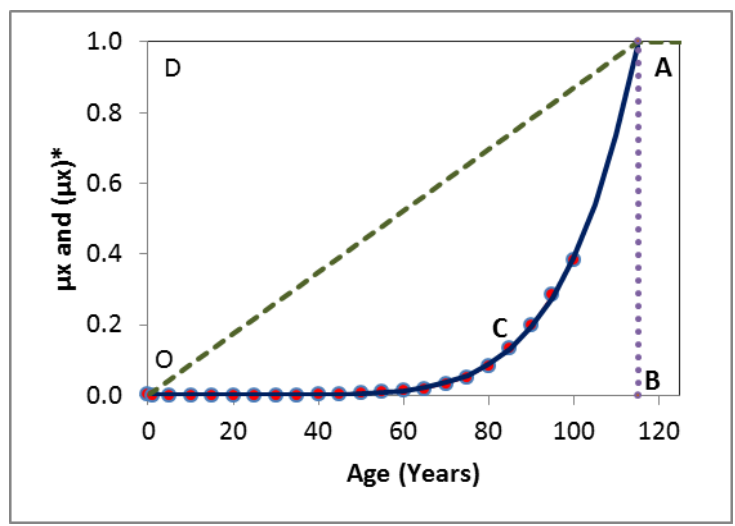

Fig. 1. The mortality diagram

The main task is to find the area $E_x$ under the curve OCABO in the mortality diagram (see Figure 1) which is a measure of the mortality effect. This is done by estimating the integral

$$E_x = \int_0^T \left(\frac{x}{T}\right)^b dx = \frac{T}{(b+1)} \left(\frac{x}{T}\right)^b$$

The resulting value for Ex in the interval [0, T] is given by the simple form:

$$E_{mortality} = \frac{T}{(b+1)}$$

It is clear that the total information for the mortality is the area provided under the curve $\mu_x$ and the horizontal axis. The total area $E_{total}$ of the healthy and mortality part of the life span is nothing else but the area included into the rectangle of length T and height 1 that is $E_{total}=T$. The health area is given by

$$E_{health} = T - E_{mortality} = T - \frac{T}{(b+1)} = \frac{bT}{b+1}$$

Then a very simple relation arises for the fraction $E_{health}/E_{mortality}$ that is

$$\frac{E_{health}}{E_{mortality}} = b \qquad (2)$$

This is the simplest indicator for the loss of health status of a population. As we have estimated by another method it is more close to the severe disability causes indicator.

The relation $E_{total}/E_{mortality}$ provides another interesting indicator of the form:

$$\frac{E_{total}}{E_{mortality}} = b + 1$$

This indicator is more appropriate for the severe and moderate disability causes indicator (It is compatible with our estimates using the health state approach). It provides larger values for the disability measures as the $E_{total}$ is larger or the $E_{mortality}$ area is smaller by means that as we live longer the disability period becomes larger.

This method suggests a simple but yet interesting tool for classification of various countries and populations, for the loss of healthy life years. A correction multiplier $\lambda$ should be added for specific situations so that the estimator of the loss of healthy life years should be of the form:

$$LHLY = \frac{E_{total}}{E_{mortality}} = \lambda(b+1)$$

However, for comparisons between countries it is sufficient to select $\lambda$=1. Even more the selection of $\lambda$=1 is appropriate when we would like to develop a quantitative measure for the LHLY without introducing the public opinion for the health status and the estimates for the cause of diseases and other disability measures. From another point of view the influence of the health status of the society to the public opinions related to health may cause differences in the values for LHLY estimated with the HALE method thus a value for $\lambda$ larger or smaller than unity is needed. By means that we will have to measure not exactly the health status but the public opinion related to the health status, the latter leading in a variety of health estimates in connection to socioeconomic and political situation along with crucial health information from the mass media. Both measures, the standard measure with $\lambda$=1 and the flexible one with $\lambda$ different from 1 could be useful for decision makes and health policy administrators and governmental planners.

To our great surprise our model by selecting $\lambda$=1 provided results very close to those provided by WHO as it is presented in the following Tables and in other applications. It is clear that we have found an interesting estimator for the loss of healthy life years.

Our idea to find the loss of healthy life years as a fraction of surfaces in a mortality diagram was proven to be quite important for expressing the health state measures. A more detailed method based on the health state stochastic theory is presented in the book on The Health State Function of a Population and related publications (see Skiadas and Skiadas 2010, 2012, 2015) where more health estimators are found.

### 2.2.1. Application details

As our method needs life table data we prefer to use full life tables when available. The Human Mortality Database is preferred for a number of countries providing full life tables. However, only a small part of the world countries are included and thus we also use the abridged life tables provided by the World Health Organization. The new abridged life tables from WHO including data from 0 to 100 years provide good results when applying our method. Instead the previous life tables (0 to 85 years) are not easily applied. It could be possible to use these life tables by expanding from 85 to 100 years. For both the abridged and the full life table data we have developed the appropriate models and estimation programs in Excel thus make it easy to use.

### 2.2.2. Stability of the coefficients of the Simple Model

Here we discuss some important issues regarding the application of the simple model proposed by equation (1). To apply this model to data we use a non-linear regression analysis technique by using a Levenberg-Marquardt algorithm. The data are obtained from the WHO database providing abridged life tables of the 0-100 years form. The important part of the model is the parameter *b* expressing the loss of healthy life years. Even more *b* can express the curvature of mortality function $\mu_x$. Applying the model to data we need a measure for the selection of the most appropriate value for *b*.

### 2.2.3. When *b* should be accepted

The simpler is to find if *b* follows a systematic change versus age. We start by selecting all the *n* data points ($m_0$, $m_1$,…, $m_n$) for $\mu_x$ to find *b* and then we select n-1, n-2,…, n-m for a sufficient number of m<n. As is presented in Figure 2 the parameter *b* follows a systematic change. The example is for USA males and females the year 2000 and the data are from the full life tables of the Human Mortality Database. As it is expected *b* is larger for females than for males. In both cases a distinct maximum value in a specific year of age appears. Accordingly a specific minimum appears for the other not so important parameter *T* (see Figure 3). It is clear that only the specific maximum value for *b* should be selected. Even more the estimates for the maximum *b* account for a local minimum for the first difference *dx'* of *dx* provided from the life table. Next Figure 4 illustrates this case for USA males the year 2000 along with a fit curve from our model SK-6. The maximum *b* is at 94 years for males and females the same as for the minimum of the first difference corresponding to the right inflection point of the death curve *dx*. Table I includes the parameter estimates for *b* and *T* the year 2000 for USA males and females.

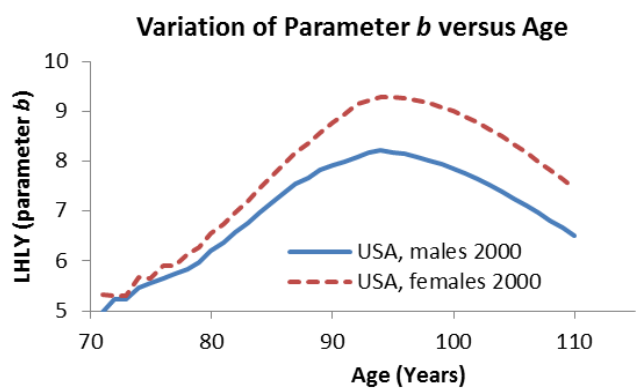

Fig. 2. Development of the health parameter

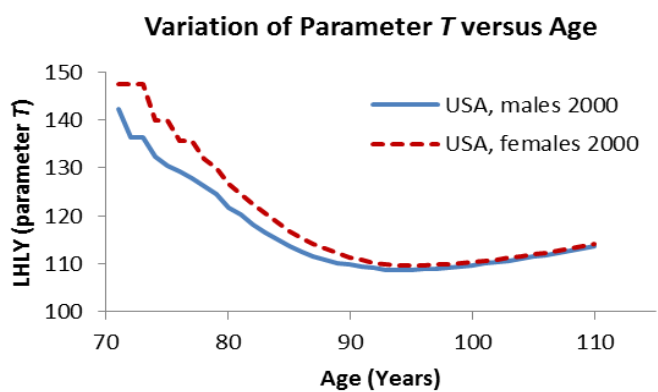

Fig. 3. Development of *T* parameter

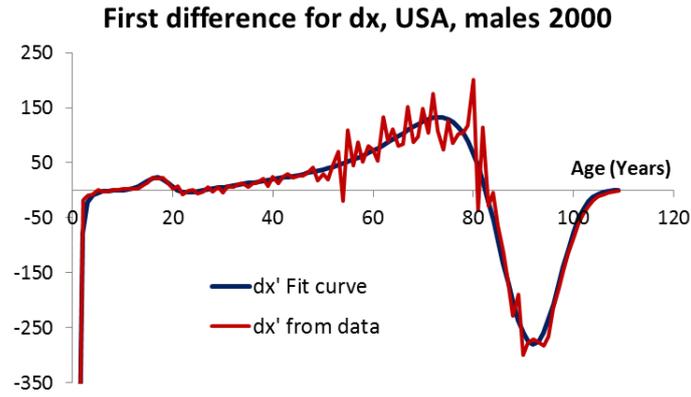

Fig. 4. First difference (derivative) of *dx* versus age

TABLE I

Parameter estimates for the model (USA, 2000)

| Age | Females | | Males | | Age | Females | | Males | |
|---|---|---|---|---|---|---|---|---|---|
| Years | b | T | b | T | Years | b | T | b | T |
| 71 | 5.318 | 147.5 | 4.975 | 142.3 | 91 | 8.942 | 110.7 | 7.992 | 109.4 |
| 72 | 5.308 | 147.5 | 5.244 | 136.4 | 92 | 9.143 | 110.0 | 8.081 | 109.1 |
| 73 | 5.296 | 147.5 | 5.231 | 136.4 | 93 | 9.224 | 109.8 | 8.173 | 108.8 |
| 74 | 5.663 | 140.0 | 5.459 | 132.3 | 94 | **9.291** | **109.6** | **8.218** | **108.6** |
| 75 | 5.649 | 140.0 | 5.559 | 130.5 | 95 | 9.286 | 109.6 | 8.189 | 108.7 |
| 76 | 5.905 | 135.6 | 5.642 | 129.2 | 96 | 9.263 | 109.6 | 8.148 | 108.8 |
| 77 | 5.896 | 135.6 | 5.736 | 127.8 | 97 | 9.224 | 109.7 | 8.094 | 109.0 |
| 78 | 6.146 | 131.9 | 5.844 | 126.3 | 98 | 9.167 | 109.9 | 8.027 | 109.2 |
| 79 | 6.280 | 130.1 | 5.981 | 124.5 | 99 | 9.093 | 110.1 | 7.947 | 109.4 |
| 80 | 6.551 | 126.8 | 6.214 | 121.8 | 100 | 9.002 | 110.3 | 7.856 | 109.7 |
| 81 | 6.748 | 124.6 | 6.368 | 120.2 | 101 | 8.896 | 110.6 | 7.754 | 110.0 |
| 82 | 6.972 | 122.5 | 6.587 | 118.2 | 102 | 8.775 | 110.8 | 7.642 | 110.3 |
| 83 | 7.209 | 120.4 | 6.774 | 116.6 | 103 | 8.641 | 111.2 | 7.521 | 110.7 |
| 84 | 7.453 | 118.5 | 6.981 | 115.0 | 104 | 8.495 | 111.5 | 7.391 | 111.0 |
| 85 | 7.710 | 116.8 | 7.186 | 113.6 | 105 | 8.339 | 111.9 | 7.255 | 111.4 |
| 86 | 7.947 | 115.3 | 7.378 | 112.5 | 106 | 8.173 | 112.3 | 7.114 | 111.8 |
| 87 | 8.185 | 114.0 | 7.546 | 111.5 | 107 | 8.000 | 112.7 | 6.967 | 112.3 |
| 88 | 8.369 | 113.1 | 7.665 | 110.9 | 108 | 7.822 | 113.1 | 6.818 | 112.7 |
| 89 | 8.579 | 112.2 | 7.826 | 110.1 | 109 | 7.638 | 113.5 | 6.666 | 113.2 |
| 90 | 8.778 | 111.3 | 7.916 | 109.8 | 110 | 7.452 | 114.0 | 6.512 | 113.6 |

## 2.3. Estimation without a model (Direct estimation)

As the needed data sets in the form of $m_x$ or $q_x$ data are provided from the life tables we have developed a method of direct estimation of the loss of healthy life year estimators directly from the life table by expanding the life table to the right.

$$b + 1 = \frac{E_{total}}{E_{mortality}} = \frac{xm_x}{\sum_0^x m_x}$$

$$b = \frac{E_{health}}{E_{mortality}} = \frac{xm_x - \sum_0^x m_x}{\sum_0^x m_x} = \frac{xm_x}{\sum_0^x m_x} - 1 \tag{3}$$

The only needed is to estimate the above fraction from the life table data. A similar indicator results by selecting the *qx* data from the life table and using the:

$$b + 1 = \frac{E_{total}}{E_{mortality}} = \frac{xq_x}{\sum_0^x q_x}$$

$$b = \frac{E_{health}}{E_{mortality}} = \frac{xq_x}{\sum_0^x q_x} - 1$$

In both cases the results are similar as it is presented in the following Figure 5 (A and B). The estimates from $m_x$ are slightly larger than from $q_x$. In both cases the *b* estimators growth to a maximum at old ages and then decline. The selected *b* or *b*+1 indicator for the life years lost from birth are those of the maximum value. A smoothing technique averaging over 5 years estimators is used to avoid sharp fluctuations in the maximum range area for the direct method. For the Model method a simple 3 point averaging gives good results. The maximum HLYL for the direct estimation is 9.84 for $m_x$ and 9.26 for $q_x$. For the Model estimation with $m_x$ data the related HLYL is 10.0. As we have estimated for other cases both the estimation of the *b* indicator by this direct method and the method by using a model give similar results.

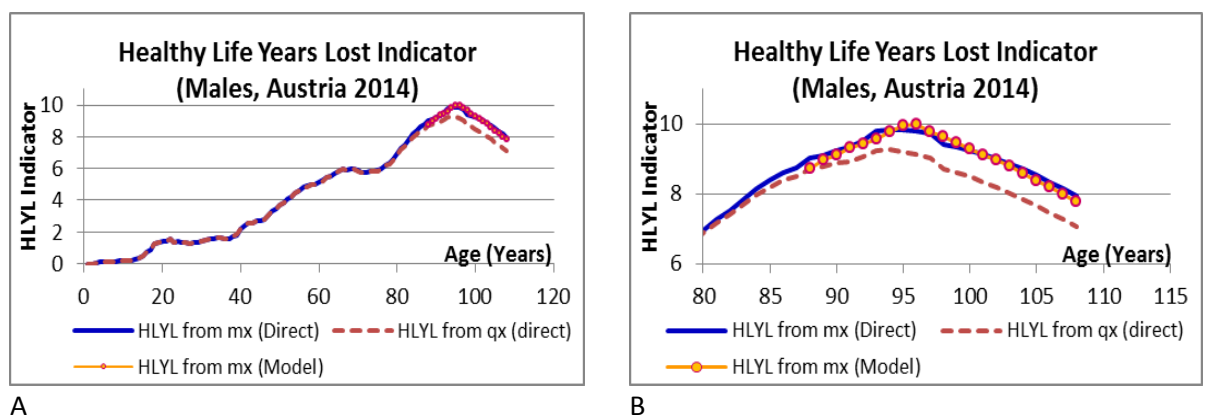

A                                       B

Fig. 5. Estimation of the HLYL indicator (*b*) by the direct method and by the simple model (Full results A and expanded around the maximum B)

## 2.4. More details: The Gompertz and the Weibull Distributions

It should be noted that a more convenient Gompertz (1825) model form is provided by Jacques F. Carriere (1992) in the form $\mu_x = Bc^x$, where *B* and *c* are parameters. This is close to our simple model selected.

However, we have also selected and applied the following form for the probability density function of the Gompertz model:

$$f_x = e^{-k+bx-e^{-l+bx}} \qquad (4)$$

The characteristic parameter expressing the loss of healthy life years is the parameter *l*. this is also demonstrated by observing the cumulative distribution function of the form:

$$F_x = e^{-e^{-l+bx}}$$

The related survival function is

$$S_x = 1 - e^{-e^{-l+bx}}$$

The probability density function is:

$$f_x = be^{-l+bx}e^{-e^{-l+bx}}$$

And the hazard function is

$$h(x) = \frac{f_x}{F_x} = be^{-l+bx} = e^{\ln(b)-l+bx} = e^{-k+bx}$$

Thus explaining the above Gompertz form selected (*k=l-*ln(*b*)).

The selected value for the estimation of the healthy life years lost is provided by the parameter *l*.

In the same paper Carriere suggests the use the Weibull model. This model has density function (*b* and *T* are parameters):

$$f_x = \frac{b}{T}\left(\frac{x}{T}\right)^{b-1} e^{-\left(\frac{x}{T}\right)^b} \qquad (5)$$

The Weibull model provides an important form for the hazard function:

$$h(x) = \frac{b}{T}\left(\frac{x}{T}\right)^{b-1}$$

Even more the cumulative hazard is given by:

$$H(x) = \left(\frac{x}{T}\right)^b$$

Another important point is that the Cumulative Hazard provided by the Weibull model is precisely the form for the simple model presented earlier and the parameter *b* expresses the healthy life years lost.

**3. The Health State Models**

**3.1. The Health State Distribution**

Although the health state models are introduced from 1995 (see Janssen and Skiadas, 2015 and more publications from Skiadas 2007 and Skiadas and Skiadas 2010, 2012, 2014) few applications appear. The main reason is due to the very laborious first exit time stochastic theory needed and that it is assumed that the use of the Gompertz and the Weibull models along with the related extensions give enough tools for the practical applications. This is not correct as the first exit time stochastic models are produced by using one of the most elegant and accurate methodology to model the health-death process as it is demonstrated in the following. The probability distribution of the general health state model is of the form:

$$f_x = \frac{|H_x - xH'_x|}{\sigma\sqrt{2\pi x^3}} e^{-\frac{H_x^2}{2x\sigma^2}}$$

For the main applications in Demography we can set σ = 1 reducing to the simpler from:

$$f_x = \frac{|H_x - xH'_x|}{\sqrt{2\pi x^3}} e^{-\frac{H_x^2}{2x}} \tag{6}$$

While the simpler form arises for the following health state function

$$H(x) = l - (bx)^c \tag{7}$$

That is

$$f_x = \frac{|l + (c-1)(bx)^c|}{\sqrt{2\pi x^3}} e^{-\frac{(l-(bx)^c)^2}{2x}} \tag{8}$$

The simpler model of this form arises when c = 1 and it is the so-called Inverse Gaussian expressing the probability density function for the first exit time of a linearly decaying process:

$$f_x = \frac{|l|}{\sqrt{2\pi x^3}} e^{-\frac{(l-bx)^2}{2x}} \tag{9}$$

Applications of this or similar type forms be can found in Ting Lee and Whitmore (2006) and in Weitz and Fraser (2001).

The last model as right skewed cannot express the human death process expressed by a highly left skewed probability density function. Instead the previous 4-parameter model is applied very successfully. Even more this form is very flexible providing very good fitting in the case of high levels of infant mortality, as it was the case for time periods some decades ago and also for nowadays when infant mortality is relatively low. Two different options arise for the model. That corresponding to the health state estimation with the parameter *l* expressing the high level of the health state and represented with the figures 6A and 6C and another form with low levels for the parameter *l* expressing the Infant Mortality (see the figures 6B and 6D). In the latter case the form of the density function is:

$$f_x = \frac{2|l + (c-1)(bx)^c|}{\sqrt{2\pi x^3}} e^{-\frac{(l-(bx)^c)^2}{2x}} \tag{10}$$

When the parameter *l* is very small a 2-parameter model termed here as the Half-Inverse Gaussian distribution results:

$$f_x = \frac{2(c-1)(bx)^c}{\sqrt{2\pi x^3}} e^{-\frac{(bx)^{2c}}{2x}} \tag{11}$$

The name arises from the similarity of this form with the Half-Normal distribution.

TABLE II

| Year | 2010 | 2010 | 2010 | 1950 | 1950 | 1950 |
|---|---|---|---|---|---|---|
| Parameter/$R^2$ | Health State | IM Model | 2-Parameter | Health State | IM Model | 2-Parameter |
| $c$ | 5.28 | 7.91 | 7.91 | 4.18 | 6.26 | 6.27 |
| $b$ | 0.0192 | 0.0148 | 0.0148 | 0.0239 | 0.0173 | 0.0173 |
| $l$ | 13.84 | 0.0066 | - | 13.05 | 0.0314 | - |
| $R^2$ | 0.993 | 0.995 | 0.993 | 0.920 | 0.990 | 0.927 |

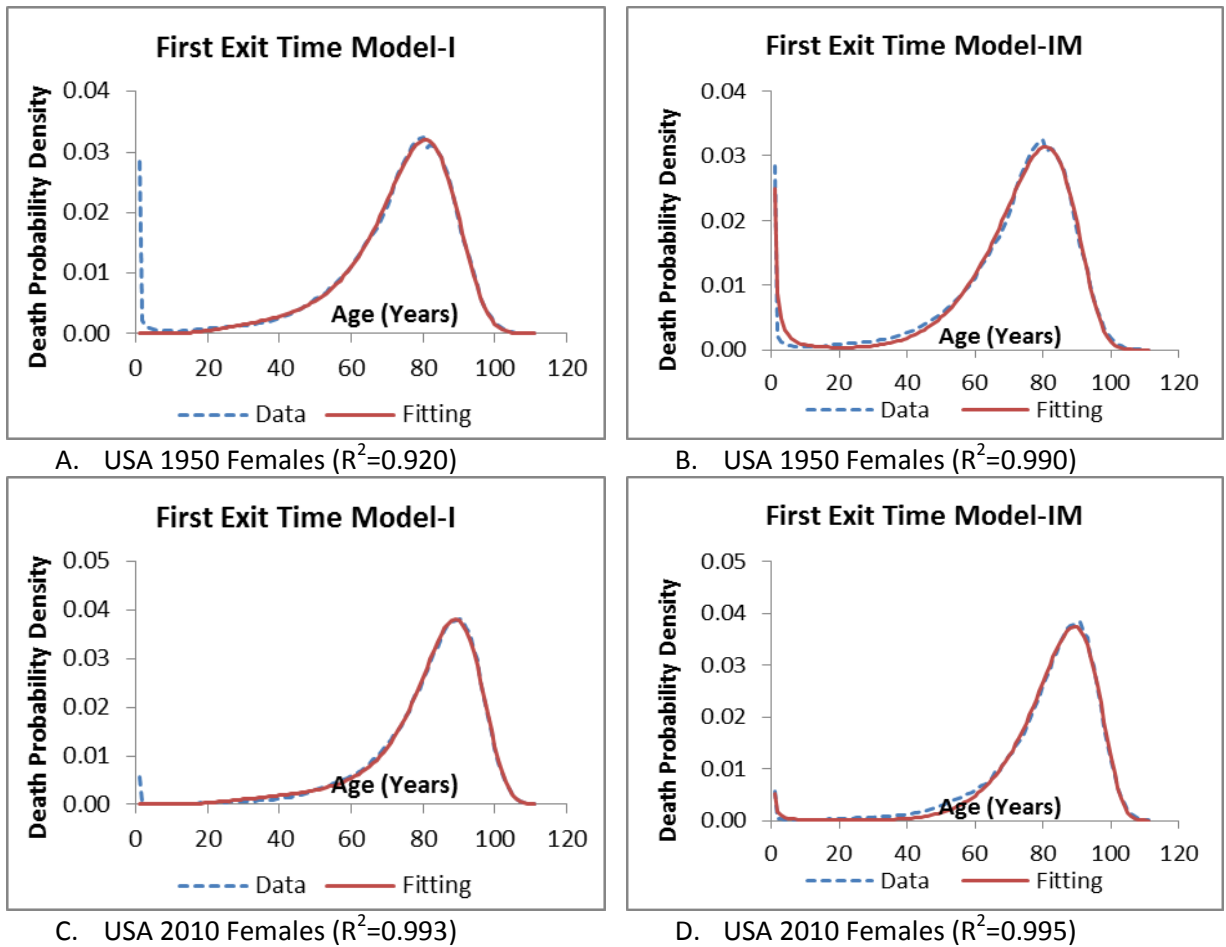

A. USA 1950 Females ($R^2$=0.920)  B. USA 1950 Females ($R^2$=0.990)
C. USA 2010 Females ($R^2$=0.993)  D. USA 2010 Females ($R^2$=0.995)

Fig. 6. The First Exit Time Model-IM including the Infant Mortality applied in USA death probability density for females the years 1950 and 2010.

The advantage of the proposed half-inverse Gaussian or IM-Model for the infant mortality modeling is obvious in the case of the application in USA females in 1950. The IM-Model provides a fairly well $R^2$=0.990 instead of $R^2$=0.920 for the Health State Model which provides similar results with the 2-parameter model (see the Table II). The resulting $R^2$ for the year 2010 in USA females are similar as the infant mortality is relatively small (see figures 6C and 6D and Table II).

**3.2. An Important Extension: The simplest IM-Model**

Christel Jennen (1985) suggested a second order approximation to improve the previous model with the first order approximation form:

$$f_x = \frac{|H_x - xH'_x|}{\sqrt{2\pi x^3}} e^{-\frac{H_x^2}{2x}}$$

However, we propose and apply here a simpler form adequate for the applications in demography data:

$$f_x = \left(\frac{2}{\sqrt{2\pi}}\right)\left(\frac{|H_x - xH'_x|}{\sqrt{x^3}} + \frac{k\sqrt{x^3}H''_x}{2|H_x - xH'_x|}\right) e^{-\frac{H_x^2}{2x}} \quad (12)$$

The parameter *k* expresses the level of the influence of the second order correction term. When *k*=0 the last equation form reduces to the first order approximation. The next step is to use the expression $H(x) = l - (bx)^c$ presented earlier for *H*(x) to find the advanced form of IM-model:

$$f_x = \left(\frac{2}{\sqrt{2\pi}}\right)\left(\frac{|l+(c-1)(bx)^c|}{\sqrt{x^3}} + \frac{k\sqrt{x^3}c(c-1)b^c x^{(c-2)}}{2|l+(c-1)(bx)^c|}\right) e^{-\frac{(l-(bx)^c)^2}{2x}} \quad (13)$$

This is the simpler 4-parameter model providing quite well fitting for the logarithm of the force of mortality, providing not only good estimates for the infant mortality but also very good estimates for all the period of the life time for males and females as is illustrated in Figures 7A-7F. We have thus demonstrated that the model proposed in 1995 and the new versions and advanced forms provided in several publications and in this paper, approach fairly well the mortality data sets provided by the bureau of the census and statistical agencies. This is important in order to straighten the findings when applying the first exit time theory to life table data.

**3.3. The Health State Function and the relative impact on mortality**

Considering the high importance of the proposed model and the related indicator for the verification of the GBD results we proceed in the introduction of a second method based on the health state of the population instead of the previous one which was based on mortality. This model was proposed earlier (see Skiadas and Skiadas, 2010, 2012, 2013, 2014). These works were based on an earlier publication modeling the health state of a population via a first exit time stochastic methodology. Here we develop a special application adapted to WHO data provided as abridged life tables (0 to 100 with 5 year periods). First we expand the abridged life table to full and then we estimate the health indicators and finally the loss of healthy life year indicators.

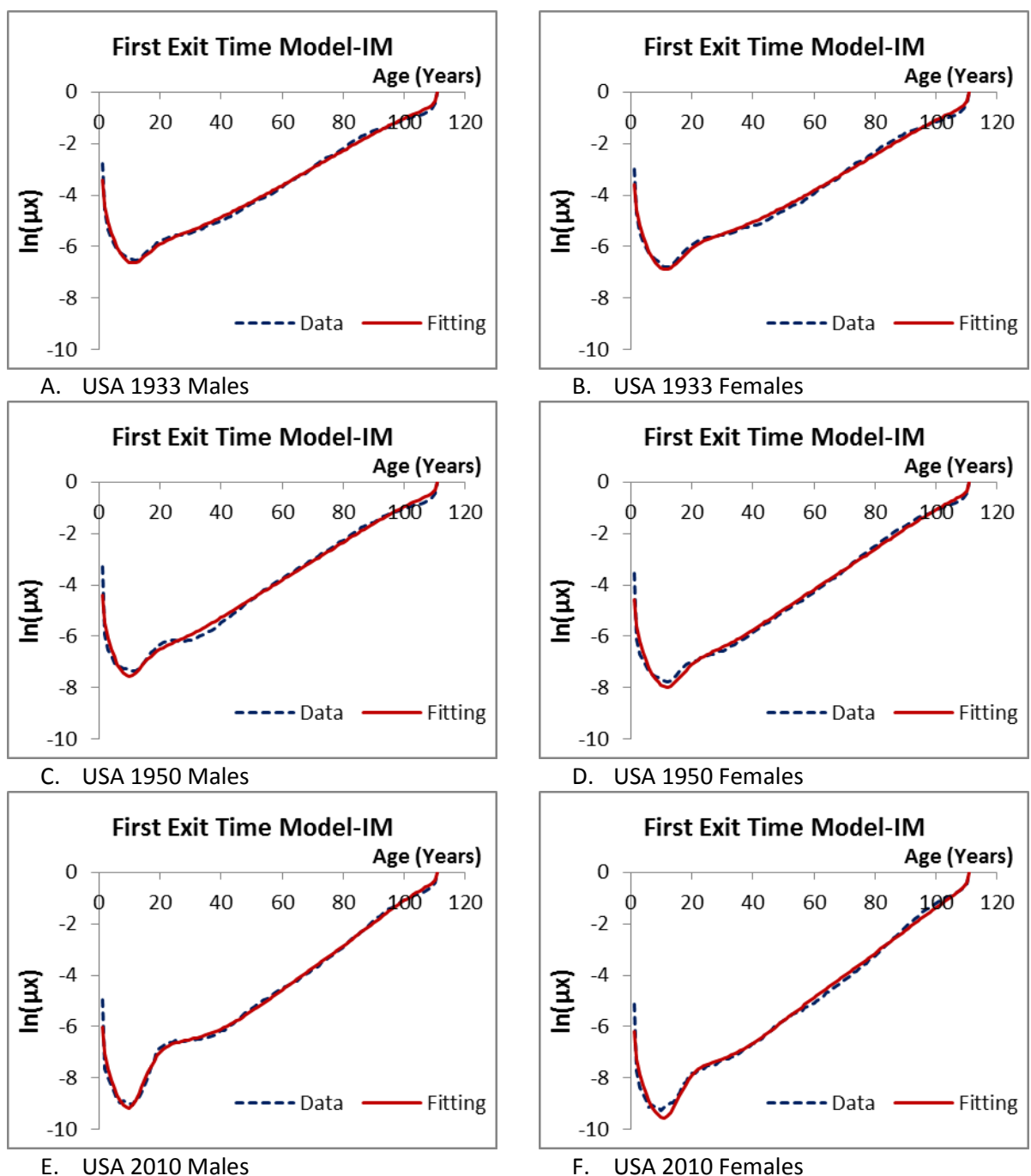

Fig. 7. The First Exit Time Model-IM including the Infant Mortality applied in the USA force of mortality data in logarithmic form for males and females the years 1933, 1950 and 2010.

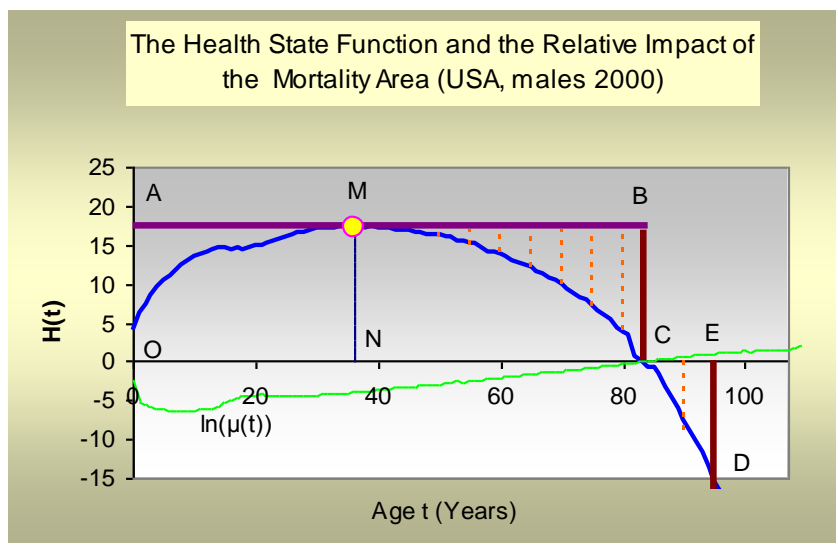

Fig. 8. The impact of the mortality area to health state

By observing the above graph (Figure 8) we can immediately see that the area between the health state curve and the horizontal axis (OMCO) represents the total health dynamics (THD) of the population. Of particular importance is also the area of the health rectangle (OABC) which includes the health state curve. This rectangle is divided in two rectangular parts the smaller (OAMN) indicating the first part of the human life until reaching the point M at the highest level of health state (usually the maximum is between 30 to 45 years) and the second part (NMBC) characterized by the gradual deterioration of the human organism until the zero level of the health state. This zero point health age C is associated with the maximum death rate. After this point the health state level appears as negative in the graph and characterizes a part of the human life totally unstable with high mortality; this is also indicated by a positively increasing form of the logarithm of the force of mortality $\ln(\mu_x)$.

We call the second rectangle NMBC as the ***deterioration rectangle***. Instead the first rectangle OAMN is here called as the ***development rectangle***. For both cases we can find the relative impact of the area inside each rectangle but outside the health state area to the overall health state. In this study we analyze the relative impact of the ***deterioration area*** MBCM indicated by dashed lines in the ***deterioration rectangle***. It should be noted that if no-deterioration mechanism was present or the repairing mechanism was perfect the health state should continue following the straight line AMB parallel to the X-axis at the level of the maximum health state. The smaller the deterioration area related to the health state area, the higher the healthy life of the population. This comparison can be done by estimating the related areas and making a simple division.

However, when trying to expand the human life further than the limits set by the deterioration mechanisms the percentage of the non-healthy life years becomes higher. This means that we need to divide the total rectangle area by that of the deterioration area to find an estimate for the "lost healthy life years". It is clear that if we don't correct the deterioration mechanisms the loss of healthy years will become higher as the expectation of life becomes larger. This is already observed in the estimates of the World Health Organization (WHO) in the World Health Report for 2000 where the lost healthy years for females are higher than the corresponding values for males. The females show higher life expectancy than males but also higher values for the lost healthy years. The proposed "loss of healthy life years" indicator is given by:

$$LHLY_1 = \lambda \frac{OABC}{THD_{ideal}} \cdot \frac{THD_{ideal}}{MBCM} = \lambda \frac{OABC}{MBCM}$$

Where $THD_{ideal}$ is ideal total health dynamics of the population and the parameter $\lambda$ expresses years and should be estimated according to the specific case. For comparing the related results in various countries we can set $\lambda=1$. When OABC approaches the $THD_{ideal}$ as is the case of several countries in nowadays the loss of healthy life years indicator LHLY can be expressed by other forms.

Another point is the use of the (ECD) area in improving forecasts especially when using the 5-year life tables as is the case of the data for all the WHO Countries. In this case the expanded loss of healthy life years indicator LHLY will take the following two forms:

$$LHLY_2 = \lambda \frac{OMCO + ECD}{MBCM}$$

$$LHLY_3 = \lambda \frac{OABC + ECD}{MBCM}$$

It is clear that the last form will give higher values than the previous one. The following scheme applies: $LHLY_1 < LHLY_2 < LHLY_3$. It remains to explore the forecasting ability of the three forms of the "loss of healthy life years" indicator by applying LHLY to life tables provided by WHO or by the Human Mortality Database or by other sources.

As for the previous case here important is the loss of health state area MBCM whereas the total area including the healthy and non-healthy part is included in OABC+ECD.

$$LHLY_3 = \lambda \frac{OABC + ECD}{MBCM} \tag{14}$$

Details and applications are included in the book on "The Health State Function of a Population", the supplement of this book and other publications (see Skiadas and Skiadas 2010, 2012, 2013, 2016). It is important that we can explore the health state of a population by using the mortality approach with the Simple Model proposed herewith and the health state function approach as well. The latter method provides many important health measures than the simple model.

**TABLE III**

| Sex/Region | Healthy Life Expectancy at Birth ||||||  Life Expectancy at Birth (LE) ||||
|---|---|---|---|---|---|---|---|---|---|---|
| | 2000 ||| 2012 ||| 2000 || 2012 ||
| | WHO (HALE) | Mortality Model | HSM Model | WHO (HALE) | Mortality Model | HSM Model | WHO | Mortality Model | WHO | Mortality Model |
| **Both sexes combined** | | | | | | | | | | |
| World | 58.0 | 58.4 | **58.2** | 61.7 | 62.5 | **61.9** | 66.2 | 66.2 | 70.3 | 70.3 |
| High income countries | 67.3 | **67.1** | 67.0 | 69.8 | **69.6** | 69.2 | 76.0 | 76.0 | 78.9 | 78.9 |
| African Region | 43.1 | **42.8** | 42.8 | 49.6 | 49.9 | **49.6** | 50.2 | 50.2 | 57.7 | 57.7 |
| Region of the Americas | 64.9 | 65.7 | **65.4** | 67.1 | 67.7 | **67.2** | 73.9 | 73.9 | 76.4 | 76.3 |
| Eastern Mediterranean Region | 55.4 | 56.9 | **56.6** | 58.3 | 59.7 | **59.4** | 64.9 | 64.9 | 67.8 | 67.8 |
| European Region | 63.9 | **63.9** | 63.9 | 66.9 | 67.2 | **67.0** | 72.4 | 72.4 | 76.1 | 76.0 |
| South East Asian Region | 54.2 | 56.3 | **55.6** | 58.5 | 60.6 | **60.0** | 62.9 | 63.0 | 67.5 | 67.5 |
| Western Pacific Region | 64.8 | **63.9** | 64.2 | 68.1 | **67.3** | 67.5 | 72.3 | 72.3 | 75.9 | 75.9 |
| **Males** | | | | | | | | | | |
| World | 56.4 | **56.6** | 56.2 | 60.1 | 60.4 | **60.0** | 63.9 | 63.9 | 68.1 | 68.0 |
| High income countries | 64.7 | 64.1 | **64.2** | 67.5 | **67.0** | 67.0 | 72.4 | 72.3 | 75.8 | 75.7 |
| African Region | 42.4 | 41.6 | **42.3** | 48.8 | **48.6** | 48.6 | 49.0 | 49.0 | 56.3 | 56.3 |
| Region of the Americas | 62.7 | 63.1 | **62.5** | 64.9 | 65.1 | **64.6** | 70.8 | 70.8 | 73.5 | 73.5 |
| Eastern Mediterranean Region | 54.8 | 55.7 | **55.6** | 57.4 | 58.2 | **57.9** | 63.6 | 63.6 | 66.1 | 66.1 |
| European Region | 60.7 | **60.4** | 61.1 | 64.2 | **64.3** | 64.5 | 68.2 | 68.2 | 72.4 | 72.4 |
| South East Asian Region | 53.5 | 55.4 | **54.6** | 57.4 | 59.2 | **58.6** | 61.6 | 61.7 | 65.7 | 65.7 |
| Western Pacific Region | 63.0 | 61.8 | **62.0** | 66.6 | 65.2 | **65.7** | 70.0 | 70.0 | 73.9 | 73.9 |
| **Females** | | | | | | | | | | |
| World | 59.7 | 60.3 | **59.9** | 63.4 | 64.3 | **64.1** | 68.5 | 68.5 | 72.7 | 72.6 |
| High income countries | 70.0 | **69.7** | 69.6 | 72.0 | 71.8 | **72.1** | 79.6 | 79.5 | 82.0 | 81.9 |
| African Region | 43.8 | **43.8** | 43.5 | 50.4 | 51.2 | **50.5** | 51.4 | 51.4 | 59.0 | 59.1 |
| Region of the Americas | 67.2 | 68.0 | **67.8** | 69.1 | 69.9 | **69.8** | 77.0 | 76.9 | 79.3 | 79.2 |
| Eastern Mediterranean Region | 56.1 | 58.2 | **57.8** | 59.2 | 61.3 | **61.0** | 66.4 | 66.4 | 69.7 | 69.6 |
| European Region | 67.1 | 67.6 | **67.3** | 69.6 | 70.0 | **69.7** | 76.7 | 76.6 | 79.6 | 79.6 |
| South East Asian Region | 55.0 | 57.2 | **56.4** | 59.7 | 62.0 | **61.7** | 64.3 | 64.4 | 69.4 | 69.4 |
| Western Pacific Region | 66.7 | 65.7 | **66.1** | 69.8 | 68.9 | **69.1** | 74.8 | 74.8 | 78.1 | 78.0 |

## 4. Applications

### 4.1. Comparative Application for the World and World Regions

The Table III includes our estimates for the healthy life expectancy at birth for the years 2000 and 2012 by applying the proposed mortality model and the health state model (HSM), and the estimates of WHO referred as HALE and included in the WHO websites (August 2015). Our estimates for the mortality model are based on LHLY=($b$+1)=$E_{total}/E_{mortality}$.

The main finding is that our models verify the WHO (HALE) estimates based on the Global Burden of Disease Study. Our results are quite close (with less to one year difference) to the estimates for the World, the High Income Countries, the African region, the European region and Western Pacific and differ by 1-2 years for the Eastern Mediterranean region and the South East Asian region. In the last two cases the collection of data and the accuracy of the information sources may lead to high uncertainty of the related health state estimates. This is demonstrated in the provided confidence intervals for the estimates in countries of these regions in the studies by Salomon et al. (2012) and the Report of WHO (2001) for the HLE of the member states (2000). From the Salomon et al. study we have calculated a mean confidence interval of 5.5 years for males and 6.8 for females for the year 2000. We thus propose to base the future works on the system we propose and to use it to calibrate the estimates especially for the countries providing of low accuracy data.

To support future studies we have formulated an easy to use framework in Excel. The only needed is to insert data for $\mu_x$ in the related column of the program. The program estimates the life expectancy, the loss of healthy life years and the healthy life expectancy.

### 4.2. Application to USA 2008

Both the Gompertz and the Weibull health estimators (section 2.4) are calculated by the appropriate computer program. The results are compared with those of the methods proposed earlier thus providing enough evidence for a successful application. The estimates of the WHO are also included in the related table. The task to find an alternative of the WHO and other estimates for the Healthy Life Years Lost is highly supported by using a series of methods leading to similar and easily reproducible results.

Only few detailed publications appear in order to use for comparative applications. It is highly appreciated that one paper by Chang et al. for USA data for 2008 and of Yong and Saito (2009) on healthy life expectancy in Japan: 1986 – 2004, published in Demographic Research are of particular importance for our comparative applications.

A very important paper by Chang et al. (2015) appeared in the Journal of Public Health. It includes calculations of the Life Expectancy (LE) and the Healthy Life Expectancy (HLE) for the United States population the year 2008 by sex and race/ethnicity.

Our task was to find good estimates compared to the authors' results by a different methodology than the survey data collection and the Sullivan method followed. Following the above provided models and estimation techniques, three different methods are selected to estimate the HLE from only the life table data sets (mortality data). The benefits are: Simple estimation, estimates for all the period where mortality data exist, comparison with existing estimates from other methods, fix the weights needed for other measurements, provide a simple methodology for health decision makers to organize future plans. As an example the estimates for USA (1950) are LE (68.0 years) and HLE (61.3 years).

We have calculated similar results for the first part of the Table 1 of the Chang et al. paper related to all races and for both sexes, male and female (see our Table IV). A power model, a Weibull model and a Gompertz model are used to first estimate the Healthy Life Years Lost (HLYL) and then find the Healthy Life Expectancy (HLE) from the simple relation HLE=LE-HLYL. We have verified that our results are within the provided "plausibility range" suggested by Chang et al. and very close to their estimates for all the life period.

Instead for the second part of Table 1 of the authors our estimates differ considerably from the related figures provided. Specifically for "Hispanic" we have estimated 8.9 HLYL corresponding to 72.1 HLE instead of 13.4 (67.6 HLE) of the authors and for "Non-Hispanic black" we estimated 7.0 HLYL (66.7 HLE) instead of 12.3 (61.4 HLE) of the authors. The other estimates for "Non-Hispanic white" 7.9 HLYL (70.5 HLE) of the authors is in agreement with our estimates of 8.3 HLYL (70.1 HLE).

It is important to clarify the cause of the differences as the perfect agreement between both methods will straighten the HLE estimators.

We have provided the related estimation programs in the webpage http://www.smtda.net/demographics2016.html to support comparative applications.

### 4.3. Application in Japan 1986-2004

The results from Yong and Saito (2009) on healthy life expectancy in Japan: 1986 – 2004, published in Demographic Research are used for our comparative applications. The authors applied the Sullivan method to data collected from a large national survey in Japan. The number of responders and the methodology applied assures relatively good results adequate for a comparative study. Especially the part of the study related to *not so poor and poor* state of health was selected for our comparisons. This is because our theory presented above suggests the estimation of the health state as a fraction of areas related to mortality and health as it is presented in Figure 1 and the related theory. The impact on the public opinion regarding the health state is due to the $E_{health}$, the $E_{mortality}$ and the total Health-Mortality area $E_{total}$. The impact is expressed by the exponent $b$ or $b+1$ depending on the form of the social status of the society and the male-female differentiation regarding the adoption and spread of the information for health, disability and mortality.

We apply the Direct Estimation as presented above in Japan from 1947 to 2012 for $m_x$ and $q_x$ data included in the full life tables provided by the Human mortality Database (HMD) thus estimating the parameters $b$ and $b+1$ as is illustrated in Figure 9 for Japan (males). In the same Figures we also have included the Healthy Life Years Lost (HLYL) from the HALE estimates of the World Health Organization. Although the results for the years 1990, 2000a, 2010, 2012 and 2013 are within the region defined by the four curves, there are significant differences in the estimates in 2000b, 2001, 2002 and 2007 underestimating the HLYL due to improvements in the methodology and the use of new epidemiological data. In the Annex Table of the World Health Report 2001 and the related of 2002 write:

*Healthy life expectancy estimates published here are not directly comparable to those published in the World Health Report 2000, due to improvements in survey methodology and the use of new epidemiological data for some diseases. See Statistical Annex notes (pp.130–135). The figures reported in this Table along with the data collection and estimation methods have been largely developed by WHO and do not necessarily reflect official statistics of Member States. Further development in collaboration with Member States is underway for improved data collection and estimation methods (WHO 2001).*

*Healthy life expectancy estimates published here are not directly comparable to those published in The World Health Report 2001, because of improvements in survey methodology and the use of new epidemiological data for some diseases and revisions of life tables for 2000 for many Member States to take new data into account (see Statistical Annex explanatory notes). The figures reported in this Table along with the data collection and estimation methods have been largely developed by WHO and do not necessarily reflect official statistics of Member States. Further development in collaboration with Member States is under way for improved data collection and estimation methods (WHO 2002).*

Figure 10 illustrates the healthy life years lost for females in Japan following the same procedure as for males. As before significant differences appear especially for the years 2002 and 2007. Even more it is clear that the differences not following a clear trend are due to the ongoing process of the estimation team of WHO to arrive in a best estimate method. To this end the recently provided estimates for 2000a, 2012 and 2013 (presented without decimal points) for males and females are closer to the results from our methodology.

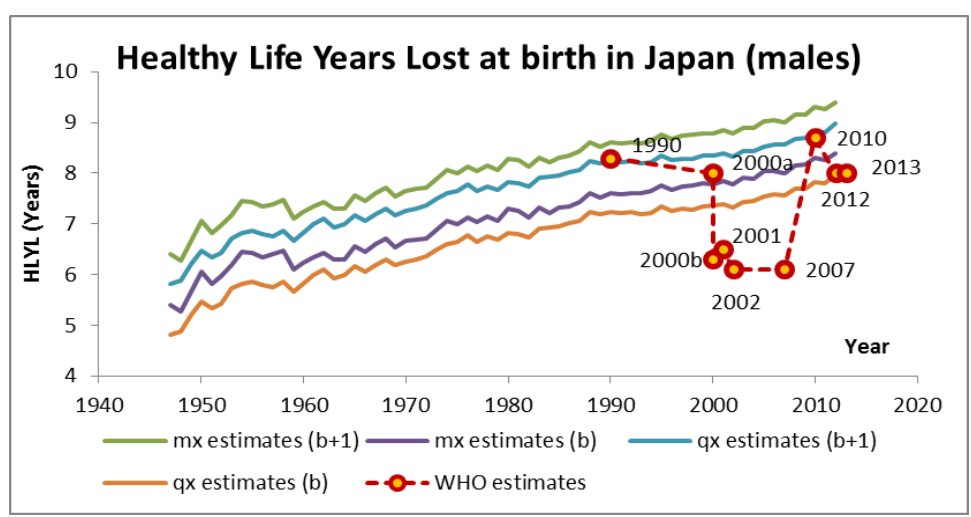

Fig. 9. Comparing the HLYL with a direct method to the WHO estimates (Japan, males)

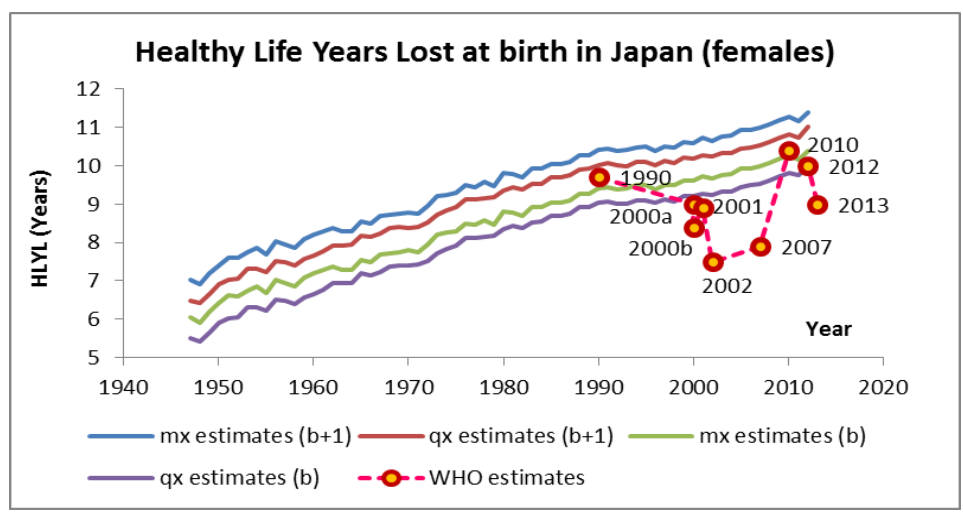

Fig. 10. Comparing the HLYL with a direct method to the WHO estimates (Japan, females)

Figures 11A – 11H illustrate the expected number of years in poor health for 25, 45, 65 and 85 years men and women following the Yong-Saito findings and the direct application results based on $m_x$ and $q_x$ estimates included in Tables VI and V (see end of the paper for these Tables and Figures 11A-11H). For both cases (men and women) the majority of the Yong-Saito estimates are within the interval suggested with our calculations. As for both men and women the Yong-Saito findings suggest a declining pattern for the healthy life years lost until 1995, followed by an increasing trend not explained by significant variations of the mortality trends or of the life expectancy, we have to explore socioeconomic factors influencing the responses to questionnaires and in a second stage the changes in the health state of a population. So, huge changes in LHLY could be expected to arise in very special morbidity cases as from the spread of epidemics. Instead the growing unemployment rate in Japan leading to a maximum in 2004 along with the slowdown of the economy and the related economic indicators can explain the relative changes in the public opinion regarding the health state. After all as the surveys cover 280,000 households and data on over 750,000 individuals were collected, the uncertainty degree should be very low. Specific sociological surveys are needed to explore the influence of socioeconomic and political factors not only to the health state but to the way the responders reply to a specific questionnaire.

## 5. Discussion and Conclusions

The GBD study critisized by Williams (see Murray et al. 2000) whereas many comments from people from social sciences and philosophy refer to the impossibility to define health and, as a consequence, to measure it. The main problem is that we cannot have flexibility in finding an estimate of health the way we do with other measures of the human organism and related activities. So far if we measure health by collecting surveys it is clear that the uncertainty is relatively high. Even more if we decide for an accepted health state estimate (see Sanders, 1964 and related studies during 60's and 70's) it remains the problem of accepting a *unit of measure*. The quantitative methods we propose overcome many of the objections posed. That we have achieved is to propose and apply several quantitative methods and techniques leading to estimates of the healthy life years lost, that more than to be close to the WHO results, provide enough evidence for estimating and quantifying the health state of a population.

Table IV. Comparing Chang *et al.* estimates

| | Life Expectancy (LE) Healthy Life Expectancy (HLE) and Healthy Life Years Lost (HLYL) by sex for the United States Population, 2008* | | | | | | | | | | | | | | | | | | | | | | | | | | | | | | |
|---|---|---|---|---|---|---|---|---|---|---|---|---|---|---|---|---|---|---|---|---|---|---|---|---|---|---|---|---|---|---|---|
| Age | Both sexes | | | | | | | | | | Male | | | | | | | | | | | Female | | | | | | | | | |
| | LE | HLE | | | | Plausibility | HLYL | | | | LE | HLE | | | | Plausibility | HLYL | | | | LE | HLE | | | | Plausibility | HLYL | | | |
| (Years) | Pub | Pub | SK | W | G | Range | Pub | SK | W | G | Pub | Pub | SK | W | G | Range | Pub | SK | W | G | Pub | Pub | SK | W | G | Range | Pub | SK | W | G |
| <1 | 78.1 | 69.3 | 68.9 | 69.7 | 69.8 | (68.4–70.3) | 8.8 | 9.2 | 8.4 | 8.3 | 75.6 | 67.6 | 66.8 | 67.7 | 67.9 | (66.8–69.2) | 8.0 | 8.8 | 7.9 | 7.7 | 80.6 | 70.9 | 70.7 | 71.2 | 71.4 | (69.8–72.6) | 9.7 | 9.9 | 9.4 | 9.2 |
| 1–4 | 77.6 | 68.7 | 68.4 | 69.2 | 69.3 | (67.9–69.8) | 8.9 | 9.2 | 8.4 | 8.3 | 75.1 | 67.1 | 66.3 | 67.2 | 67.4 | (66.1–68.5) | 8.0 | 8.8 | 7.9 | 7.7 | 80.1 | 70.3 | 70.2 | 70.7 | 70.9 | (69.1–71.8) | 9.8 | 9.9 | 9.4 | 9.2 |
| 5–9 | 73.7 | 64.9 | 64.6 | 65.3 | 65.5 | (64.1–66.0) | 8.8 | 9.1 | 8.4 | 8.2 | 71.2 | 63.2 | 62.5 | 63.3 | 63.5 | (62.2–64.6) | 8.0 | 8.7 | 7.9 | 7.7 | 76.1 | 66.5 | 66.2 | 66.8 | 66.9 | (65.2–67.9) | 9.6 | 9.9 | 9.3 | 9.2 |
| 10–14 | 68.8 | 60.0 | 59.7 | 60.5 | 60.6 | (59.2–61.1) | 8.8 | 9.1 | 8.3 | 8.2 | 66.3 | 58.4 | 57.6 | 58.5 | 58.7 | (57.4–59.7) | 7.9 | 8.7 | 7.8 | 7.6 | 71.2 | 61.6 | 61.4 | 61.9 | 62.1 | (60.4–63.0) | 9.6 | 9.8 | 9.3 | 9.1 |
| 15–19 | 63.8 | 55.1 | 54.8 | 55.6 | 55.7 | (54.4–56.2) | 8.7 | 9.0 | 8.2 | 8.1 | 61.3 | 53.5 | 52.7 | 53.6 | 53.8 | (52.6–54.8) | 7.8 | 8.6 | 7.7 | 7.5 | 66.2 | 56.7 | 56.5 | 57.1 | 57.2 | (55.5–58.1) | 9.5 | 9.7 | 9.1 | 9.0 |
| 20–24 | 59.0 | 50.4 | 50.2 | 50.9 | 51.1 | (49.7–51.5) | 8.6 | 8.8 | 8.1 | 7.9 | 56.6 | 48.8 | 48.2 | 49.0 | 49.2 | (47.9–50.1) | 7.8 | 8.4 | 7.6 | 7.4 | 61.3 | 52.0 | 51.8 | 52.3 | 52.5 | (50.8–53.3) | 9.3 | 9.5 | 9.0 | 8.8 |
| 25–29 | 54.3 | 45.9 | 45.7 | 46.4 | 46.6 | (45.2–46.9) | 8.4 | 8.6 | 7.9 | 7.7 | 52.0 | 44.4 | 43.8 | 44.6 | 44.8 | (43.5–45.6) | 7.6 | 8.2 | 7.4 | 7.2 | 56.5 | 47.3 | 47.2 | 47.8 | 47.9 | (46.2–48.6) | 9.2 | 9.3 | 8.7 | 8.6 |
| 30–34 | 49.5 | 41.4 | 41.2 | 41.9 | 42.0 | (40.7–42.3) | 8.1 | 8.3 | 7.6 | 7.5 | 47.3 | 39.9 | 39.4 | 40.2 | 40.3 | (39.1–41.1) | 7.4 | 7.9 | 7.1 | 7.0 | 51.6 | 42.7 | 42.6 | 43.1 | 43.3 | (41.7–44.0) | 8.9 | 9.0 | 8.5 | 8.3 |
| 35–39 | 44.8 | 36.8 | 36.8 | 37.5 | 37.6 | (36.2–37.8) | 8.0 | 8.0 | 7.3 | 7.2 | 42.6 | 35.4 | 35.0 | 35.7 | 35.9 | (34.7–36.6) | 7.2 | 7.6 | 6.9 | 6.7 | 46.8 | 38.2 | 38.2 | 38.7 | 38.8 | (37.2–39.4) | 8.6 | 8.6 | 8.1 | 8.0 |
| 40–44 | 40.1 | 32.5 | 32.5 | 33.1 | 33.3 | (31.9–33.4) | 7.6 | 7.6 | 7.0 | 6.8 | 38.0 | 31.0 | 30.7 | 31.4 | 31.6 | (30.3–32.2) | 7.0 | 7.3 | 6.6 | 6.4 | 42.0 | 33.8 | 33.8 | 34.2 | 34.4 | (32.9–34.9) | 8.2 | 8.2 | 7.8 | 7.6 |
| 45–49 | 35.5 | 28.3 | 28.3 | 28.9 | 29.0 | (27.8–29.2) | 7.2 | 7.2 | 6.6 | 6.5 | 33.5 | 26.9 | 26.6 | 27.3 | 27.5 | (26.3–28.0) | 6.6 | 6.9 | 6.2 | 6.0 | 37.3 | 29.6 | 29.5 | 30.0 | 30.1 | (28.7–30.7) | 7.7 | 7.8 | 7.3 | 7.2 |
| 50–54 | 31.0 | 24.3 | 24.3 | 24.8 | 24.9 | (23.8–25.1) | 6.7 | 6.7 | 6.2 | 6.1 | 29.1 | 22.9 | 22.7 | 23.3 | 23.4 | (22.4–24.0) | 6.2 | 6.4 | 5.8 | 5.7 | 32.8 | 25.5 | 25.5 | 25.9 | 26.1 | (24.8–26.6) | 7.3 | 7.3 | 6.9 | 6.7 |
| 55–59 | 26.8 | 20.6 | 20.6 | 21.1 | 21.2 | (20.2–21.4) | 6.2 | 6.2 | 5.7 | 5.6 | 25.0 | 19.3 | 19.1 | 19.7 | 19.8 | (18.8–20.3) | 5.7 | 5.9 | 5.3 | 5.2 | 28.4 | 21.8 | 21.7 | 22.1 | 22.2 | (21.1–22.7) | 6.6 | 6.7 | 6.3 | 6.2 |
| 60–64 | 22.7 | 17.2 | 17.0 | 17.5 | 17.6 | (16.8–17.9) | 5.5 | 5.7 | 5.2 | 5.1 | 21.0 | 16.0 | 15.6 | 16.1 | 16.3 | (15.5–16.9) | 5.0 | 5.4 | 4.9 | 4.7 | 24.1 | 18.2 | 18.0 | 18.3 | 18.4 | (17.6–19.1) | 5.9 | 6.1 | 5.8 | 5.7 |
| 65–69 | 18.8 | 14.0 | 13.8 | 14.2 | 14.3 | (13.7–14.7) | 4.8 | 5.0 | 4.6 | 4.5 | 17.3 | 13.0 | 12.5 | 13.0 | 13.1 | (12.6–13.8) | 4.3 | 4.8 | 4.3 | 4.2 | 20.0 | 14.9 | 14.6 | 14.9 | 15.0 | (14.3–15.7) | 5.1 | 5.4 | 5.1 | 5.0 |
| 70–74 | 15.2 | 11.1 | 10.8 | 11.2 | 11.3 | (10.8–11.7) | 4.1 | 4.4 | 4.0 | 3.9 | 13.9 | 10.2 | 9.8 | 10.1 | 10.2 | (9.9–11.0) | 3.7 | 4.1 | 3.8 | 3.7 | 16.2 | 11.7 | 11.5 | 11.7 | 11.8 | (11.3–12.5) | 4.5 | 4.7 | 4.5 | 4.4 |
| 75–79 | 11.8 | 8.4 | 8.1 | 8.4 | 8.5 | (8.2–9.0) | 3.4 | 3.7 | 3.4 | 3.3 | 10.7 | 7.7 | 7.2 | 7.6 | 7.6 | (7.5–8.4) | 3.0 | 3.5 | 3.1 | 3.1 | 12.6 | 8.9 | 8.7 | 8.9 | 8.9 | (8.6–9.6) | 3.7 | 3.9 | 3.7 | 3.7 |
| 80–84 | 8.9 | 6.1 | 6.0 | 6.2 | 6.3 | (6.0–6.7) | 2.8 | 2.9 | 2.7 | 2.6 | 8.0 | 5.5 | 5.3 | 5.5 | 5.6 | (5.3–6.1) | 2.5 | 2.7 | 2.5 | 2.4 | 9.5 | 6.6 | 6.4 | 6.5 | 6.6 | (6.4–7.2) | 2.9 | 3.1 | 3.0 | 2.9 |
| 85+ | 6.4 | 4.3 | 4.3 | 4.5 | 4.5 | (4.3–4.8) | 2.1 | 2.1 | 1.9 | 1.9 | 5.7 | 3.7 | 3.7 | 3.9 | 3.9 | (3.6–4.3) | 2.0 | 2.0 | 1.8 | 1.8 | 6.8 | 4.6 | 4.6 | 4.7 | 4.7 | (4.5–5.1) | 2.2 | 2.2 | 2.1 | 2.1 |

Pub: Chang et al. estimates    SK: Skiadas model estimates    W: Weibull estimates    HLE-G: Gompertz estimates

TABLE V

| Life expectancy and healthy life expectancy for Japanese men, 1986-2004 ||||||
|---|---|---|---|---|---|---|
| Year | Life Expectancy | Expected number of years in poor health |||||
|  |  | Yong-Saito | mx(b+1) | qx(b+1) | mx(b) | qx(b) |
| At birth men |||||||
| 1986 | **75.3** |  | 8.3 | 8.0 | 7.3 | 7.0 |
| 1989 | **76.0** |  | 8.5 | 8.2 | 7.5 | 7.2 |
| 1992 | **76.1** |  | 8.6 | 8.2 | 7.6 | 7.2 |
| 1995 | **76.4** |  | 8.8 | 8.3 | 7.8 | 7.3 |
| 1998 | **77.2** |  | 8.8 | 8.3 | 7.8 | 7.3 |
| 2001 | **78.0** |  | 8.8 | 8.4 | 7.8 | 7.4 |
| 2004 | **78.6** |  | 8.9 | 8.4 | 7.9 | 7.4 |
| 25 year old men |||||||
| 1986 | **51.4** | **7.2** | 7.8 | 7.5 | 6.9 | 6.5 |
| 1989 | **52.0** | **7.2** | 8.0 | 7.6 | 7.0 | 6.7 |
| 1992 | **52.1** | **6.3** | 8.0 | 7.7 | 7.1 | 6.7 |
| 1995 | **52.3** | **5.7** | 8.2 | 7.8 | 7.2 | 6.8 |
| 1998 | **53.0** | **6.9** | 8.2 | 7.7 | 7.2 | 6.8 |
| 2001 | **53.8** | **7.7** | 8.3 | 7.8 | 7.3 | 6.9 |
| 2004 | **54.3** | **8.0** | 8.3 | 7.9 | 7.4 | 6.9 |
| 45 year old men |||||||
| 1986 | **32.4** | **5.7** | 6.5 | 6.3 | 5.8 | 5.5 |
| 1989 | **32.9** | **5.9** | 6.7 | 6.4 | 5.9 | 5.6 |
| 1992 | **33.0** | **5.0** | 6.7 | 6.4 | 6.0 | 5.7 |
| 1995 | **33.3** | **4.6** | 6.9 | 6.5 | 6.1 | 5.7 |
| 1998 | **34.0** | **5.5** | 6.9 | 6.5 | 6.1 | 5.7 |
| 2001 | **34.8** | **6.2** | 6.9 | 6.6 | 6.1 | 5.8 |
| 2004 | **35.3** | **6.5** | 7.0 | 6.6 | 6.2 | 5.8 |
| 65 year old men |||||||
| 1986 | **15.9** | **3.8** | 4.6 | 4.0 | 4.4 | 3.8 |
| 1989 | **16.2** | **3.9** | 4.7 | 4.1 | 4.5 | 3.9 |
| 1992 | **16.3** | **3.4** | 4.7 | 4.2 | 4.5 | 4.0 |
| 1995 | **16.5** | **3.1** | 4.8 | 4.3 | 4.6 | 4.0 |
| 1998 | **17.1** | **3.8** | 4.8 | 4.3 | 4.5 | 4.0 |
| 2001 | **17.8** | **4.4** | 4.9 | 4.3 | 4.6 | 4.1 |
| 2004 | **18.2** | **4.7** | 4.9 | 4.3 | 4.6 | 4.1 |
| 85 year old men |||||||
| 1986 | **4.8** | **1.4** | 1.9 | 1.7 | 1.8 | 1.6 |
| 1989 | **4.9** | **1.5** | 1.9 | 1.7 | 1.9 | 1.6 |
| 1992 | **4.9** | **1.2** | 2.0 | 1.7 | 1.9 | 1.6 |
| 1995 | **5.1** | **1.1** | 2.0 | 1.8 | 1.9 | 1.7 |
| 1998 | **5.5** | **1.6** | 2.0 | 1.8 | 1.9 | 1.7 |
| 2001 | **5.9** | **2.1** | 2.0 | 1.8 | 1.9 | 1.7 |
| 2004 | **6.1** | **2.3** | 2.0 | 1.8 | 1.9 | 1.7 |

TABLE VI

| Life expectancy and healthy life expectancy for Japanese women, 1986-2004 ||||||
| Year | Life Expectancy | Expected number of years in poor health ||||
| | | Yong-Saito | mx(b+1) | qx(b+1) | mx(b) | qx(b) |
|---|---|---|---|---|---|---|
| At birth men |||||||
| 1986 | **81.0** |  | 10.0 | 9.7 | 9.0 | 8.7 |
| 1989 | **81.8** |  | 10.3 | 9.9 | 9.3 | 8.9 |
| 1992 | **82.3** |  | 10.4 | 10.0 | 9.4 | 9.0 |
| 1995 | **82.8** |  | 10.5 | 10.1 | 9.5 | 9.1 |
| 1998 | **83.9** |  | 10.5 | 10.1 | 9.5 | 9.1 |
| 2001 | **84.9** |  | 10.7 | 10.3 | 9.7 | 9.3 |
| 2004 | **85.5** |  | 10.8 | 10.3 | 9.8 | 9.3 |
| 25 year old women |||||||
| 1986 | **56.7** | **9.8** | 9.4 | 9.1 | 8.4 | 8.1 |
| 1989 | **57.5** | **9.8** | 9.6 | 9.3 | 8.7 | 8.3 |
| 1992 | **57.9** | **8.8** | 9.7 | 9.3 | 8.8 | 8.4 |
| 1995 | **58.6** | **8.0** | 9.8 | 9.4 | 8.9 | 8.5 |
| 1998 | **59.6** | **9.8** | 9.8 | 9.4 | 8.9 | 8.5 |
| 2001 | **60.5** | **11.0** | 10.0 | 9.6 | 9.1 | 8.6 |
| 2004 | **61.1** | **11.2** | 10.1 | 9.6 | 9.1 | 8.7 |
| 45 year old women |||||||
| 1986.0 | **37.4** | **8.0** | 7.9 | 7.6 | 7.1 | 6.8 |
| 1989.0 | **38.1** | **8.2** | 8.1 | 7.8 | 7.3 | 7.0 |
| 1992.0 | **38.5** | **7.2** | 8.1 | 7.8 | 7.4 | 7.1 |
| 1995.0 | **39.1** | **6.6** | 8.2 | 7.9 | 7.4 | 7.1 |
| 1998.0 | **40.2** | **8.0** | 8.2 | 7.9 | 7.4 | 7.1 |
| 2001.0 | **41.0** | **9.0** | 8.4 | 8.0 | 7.6 | 7.3 |
| 2004.0 | **41.6** | **9.2** | 8.5 | 8.1 | 7.7 | 7.3 |
| 65 year old women |||||||
| 1986 | **19.3** | **5.2** | 5.5 | 5.3 | 5.0 | 4.8 |
| 1989 | **20.0** | **5.4** | 5.6 | 5.4 | 5.1 | 4.9 |
| 1992 | **20.3** | **4.9** | 5.7 | 5.5 | 5.1 | 4.9 |
| 1995 | **20.9** | **4.6** | 5.7 | 5.5 | 5.2 | 5.0 |
| 1998 | **22.0** | **5.6** | 5.7 | 5.5 | 5.2 | 5.0 |
| 2001 | **22.7** | **6.6** | 5.9 | 5.6 | 5.3 | 5.1 |
| 2004 | **23.3** | **6.8** | 5.9 | 5.7 | 5.4 | 5.1 |
| 85 year old women |||||||
| 1986 | **5.7** | **1.7** | 2.3 | 2.2 | 2.1 | 2.0 |
| 1989 | **6.0** | **1.8** | 2.3 | 2.3 | 2.1 | 2.0 |
| 1992 | **6.1** | **1.7** | 2.4 | 2.3 | 2.1 | 2.0 |
| 1995 | **6.7** | **1.8** | 2.4 | 2.3 | 2.2 | 2.1 |
| 1998 | **7.4** | **2.2** | 2.4 | 2.3 | 2.2 | 2.1 |
| 2001 | **7.8** | **3.1** | 2.4 | 2.3 | 2.2 | 2.1 |
| 2004 | **8.1** | **3.1** | 2.5 | 2.3 | 2.2 | 2.1 |

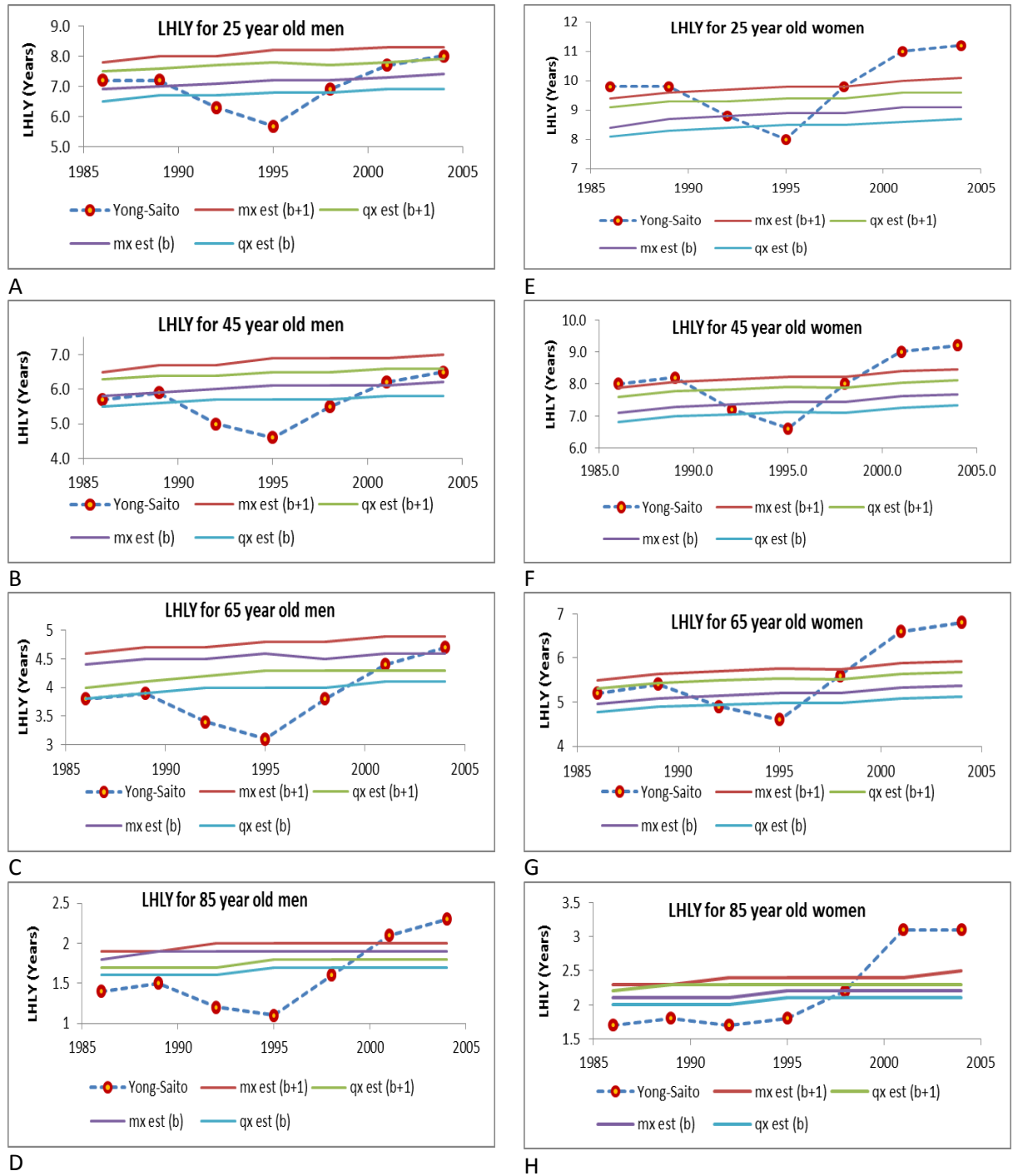

Fig. 11(A-H). Expected number of years in poor health for 25, 45, 65 and 85 years men and women (Yong-Saito findings and direct application results based on $m_x$ and $q_x$)